\documentclass[aps,showpacs, showkeys,nofootinbib,floatfix]{revtex4}

\usepackage{amssymb}
\usepackage{amsmath}
\usepackage{graphicx}
\begin{document}

\title{Time Variable Cosmological Constants from the Age of Universe}

\author{Lixin Xu\footnote{Corresponding author}}
\email{lxxu@dlut.edu.cn}
\author{Jianbo Lu}
\author{Wenbo Li}

\affiliation{Institute of Theoretical Physics, School of Physics \&
Optoelectronic Technology, Dalian University of Technology, Dalian,
116024, P. R. China}

\begin{abstract}
In this paper, time variable cosmological constant, dubbed {\it age
cosmological constant}, is investigated motivated by the fact: any
cosmological length scale and time scale can introduce a
cosmological constant or vacuum energy density into Einstein's
theory. The age cosmological constant takes the form
$\rho_{\Lambda}=3c^2M^2_P/t_{\Lambda}^2$, where $t_{\Lambda}$ is the
age of our universe or conformal time. The effective equation of
state of age cosmological constant are
$w^{eff}_{\Lambda}=-1+\frac{2}{3}\frac{\sqrt{\Omega_{\Lambda}}}{c}$
and
$w^{eff}_{\Lambda}=-1+\frac{2}{3}\frac{\sqrt{\Omega_{\Lambda}}}{c}(1+z)$
when the age of universe and conformal time are taken as the role of
cosmological time scales respectively. They are the same as the
so-called agegraphic dark energy models. However, the evolution
history are different from the agegraphic ones for their different
evolution equations.
\end{abstract}

%\pacs{Added}

 %\hfill ITP-DUT/2009-04

\maketitle

\section{Introduction}

The observation of the Supernovae of type Ia
\cite{ref:Riess98,ref:Perlmuter99} provides the evidence that the
universe is undergoing accelerated expansion. Jointing the
observations from Cosmic Background Radiation
\cite{ref:Spergel03,ref:Spergel06} and SDSS
\cite{ref:Tegmark1,ref:Tegmark2}, one concludes that the universe at
present is dominated by $70\%$ exotic component, dubbed dark energy,
which has negative pressure and push the universe to accelerated
expansion. Of course, a natural explanation to the accelerated
expansion is due to a positive tiny cosmological constant. Though,
it suffers the so-called {\it fine tuning} and {\it cosmic
coincidence} problems. However, in $2\sigma$ confidence level, it
fits the observations very well \cite{ref:Komatsu}. If the
cosmological constant is not a real constant but is time variable,
the fine tuning and cosmic coincidence problems can be removed. In
fact, this possibility was considered in the past years.

In particular, the dynamic vacuum energy density based on
holographic principle are investigated extensively
\cite{ref:holo1,ref:holo2}. According to the holographic principle,
the number of degrees of freedom in a bounded system should be
finite and has relations with the area of its boundary. By applying
the principle to cosmology, one can obtain the upper bound of the
entropy contained in the universe. For a system with size $L$ and UV
cut-off $\Lambda$ without decaying into a black hole, it is required
that the total energy in a region of size $L$ should not exceed the
mass of a black hole of the same size, thus $L^3\rho_{\Lambda} \le L
M^2_{P}$. The largest $L$ allowed is the one saturating this
inequality, thus $\rho_{\Lambda} =3c^2 M^{2}_{P} L^{-2}$, where $c$
is a numerical constant and $M_{P}$ is the reduced Planck Mass
$M^{-2}_{P}=8 \pi G$. It just means a {\it duality} between UV
cut-off and IR cut-off. The UV cut-off is related to the vacuum
energy, and IR cut-off is related to the large scale of the
universe, for example Hubble horizon, future event horizon or
particle horizon as discussed by
\cite{ref:holo1,ref:holo2,ref:Horvat1,ref:Horvat2}. The holographic
dark energy in Brans-Dicke theory is also studied in Ref.
\cite{ref:BransDicke,ref:BDH1,ref:BDH2,ref:BDH3,ref:BDH4,ref:BDH5}.

Another dark energy model with relations with holographic dark
energy, named agegraphic dark energy, was also researched
extensively recently
\cite{ref:agegraphic1,ref:agegraphic2,ref:agegraphic3}. This model
is based on the application of the well-known Heisenberg uncertainty
relation to the universe. Therefore, the energy density of metric
fluctuations of Minkowski space-time is $\rho_{\Lambda}\sim
M^2_P/t^2$, where $t$ is time or length scale. Obviously, it looks
like the holographic one. Indeed, there some relations between them
\cite{ref:agegraphic1}.

As known, for any nonzero value of the cosmological constant
$\Lambda$, a natural length scale and time scale
\begin{equation}
r_{\Lambda}=t_{\Lambda}=\sqrt{3/|\Lambda|}
\end{equation}
can be introduced into Einstein's theory. Reversely, any
cosmological length scale and time scale can introduce a
cosmological constant or vacuum energy density into Einstein's
theory. For a positive cosmological constant, one has
\begin{equation}
\Lambda(t)=\frac{3}{r^2_{\Lambda}(t)}=\frac{3}{t^2_{\Lambda}}.
\end{equation}
When a dynamic time scale is taken, a time variable cosmological
constant can be obtained. Obviously, a natural time scale is the age
of our universe. Inspired by this observation, we can consider time
variable cosmological constant from this analogue and let the
holographic principle and Heisenberg uncertainty relation alone. For
its explicit relation with the age of the universe, we dub it {\it
age cosmological constant}.

This paper is structured as follows. In Section \ref{sec:TVC}, we
give a brief review of a time variable cosmological constant. In
Section \ref{sec:age}, time variable cosmological constants--age
cosmological constants--are investigated, where the age of our
universe and conformal time are taken as the role of time scales.
Section \ref{sec:Con} are conclusions.

\section{Time Variable Cosmological Constant}\label{sec:TVC}

The Einstein equation with a cosmological constant is written as
\begin{equation}
R_{\mu\nu}-\frac{1}{2}Rg_{\mu\nu}+\Lambda g_{\mu\nu}=8\pi G
T_{\mu\nu},\label{eq:EE}
\end{equation}
where $T_{\mu\nu}$ is the energy-momentum tensor of ordinary matter
and radiation. From the Bianchi identity, one has the conservation
of the energy-momentum tensor $\nabla^{\mu}T_{\mu\nu}=0$, it follows
necessarily that $\Lambda$ is a constant. To have a time variable
cosmological constant $\Lambda=\Lambda(t)$, one can move the
cosmological constant to the right hand side of Eq. (\ref{eq:EE})
and take $\tilde{T}_{\mu\nu}=T_{\mu\nu}-\frac{\Lambda(t)}{8\pi
G}g_{\mu\nu}$ as the total energy-momentum tenor. Once again to
preserve the Bianchi identity or local energy-momentum conservation
law, $\nabla^{\mu}\tilde{T}_{\mu\nu}=0$, one has, in a spacial flat
FRW universe,
\begin{equation}
\dot{\rho}_{\Lambda}+\dot{\rho}_{m}+3H\left(1+w_{m}\right)\rho_{m}=0,\label{eq:conservation}
\end{equation}
where $\rho_{\Lambda}=M^2_{P}\Lambda(t)$ is the energy density of
time variable cosmological constant and its equation of state is
$w_{\Lambda}=-1$, and $w_{m}$ is the equation of state of ordinary
matter, for dark matter $w_m=0$. It is natural to consider
interactions between variable cosmological constant and dark matter
\cite{ref:Horvat2}, as seen from Eq. (\ref{eq:conservation}). After
introducing an interaction term $Q$, one has
\begin{eqnarray}
\dot{\rho}_{m}+3H\left(1+w_{m}\right)\rho_{m}=Q,\label{eq:rhom} \\
\dot{\rho}_{\Lambda}+3H\left(\rho_{\Lambda}+p_{\Lambda}\right)=-Q,\label{eq:rholambda}
\end{eqnarray}
and the total energy-momentum conservation equation
\begin{equation}
\dot{\rho}_{tot}+3H\left(\rho_{tot}+p_{tot}\right)=0.
\end{equation}
For a time variable cosmological constant, the equality
$\rho_{\Lambda}+p_{\Lambda}=0$ still holds. Immediately, one has the
interaction term $Q=-\dot{\rho}_{\Lambda}$ which is different from
the interactions between dark matter and dark energy considered in
the literatures \cite{ref:interaction} where a general interacting
form $Q=3b^2H\left(\rho_{m}+\rho_{\Lambda}\right)$ is put by hand.
With observation to Eq. (\ref{eq:rholambda}), the interaction term
$Q$ can be moved to the left hand side of the equation, and one has
the effective pressure of variable cosmological constant, dark
energy
\begin{equation}
\dot{\rho}_{\Lambda}+3H\left(\rho_{\Lambda}+p^{eff}_{\Lambda}\right)=0
\end{equation}
where $p^{eff}_{\Lambda}=p_{\Lambda}+\frac{Q}{3H}$ is the effective
dark energy pressure. Also, one can define the effective equation of
state of dark energy
\begin{eqnarray}
w^{eff}_{\Lambda}&=&\frac{p^{eff}_{\Lambda}}{\rho_{\Lambda}}\nonumber\\
&=&-1+\frac{Q}{3H\rho_{\Lambda}}\nonumber\\
&=&=-1-\frac{1}{3}\frac{d \ln \rho_{\Lambda}}{d\ln
a}.\label{eq:EEOS}
\end{eqnarray}
The Friedmann equation as usual can be written as, in a spacial flat
FRW universe,
\begin{equation}
H^2=\frac{1}{3M^2_P}\left(\rho_{m}+\rho_{\Lambda}\right)\label{eq:FE}.
\end{equation}

\section{Age cosmological constants}\label{sec:age}

\subsection{Age of the universe as time scale}\label{sec:ATS}

\begin{figure}[tbh]
\centering
\includegraphics[width=5in]{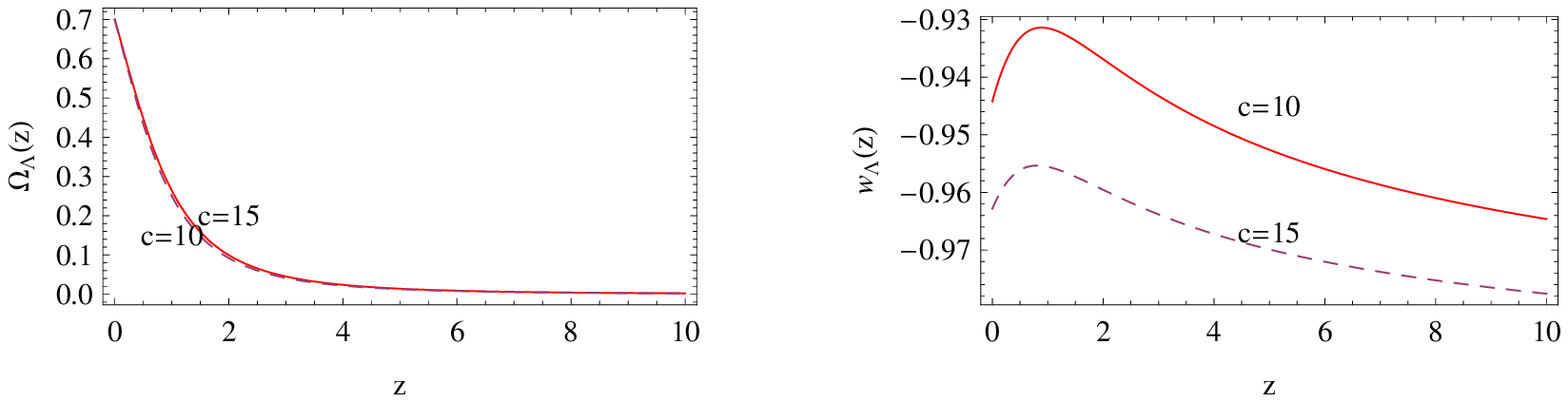}
\caption{The evolutions of dimensionless density parameter
$\Omega_{\Lambda}(z)$ (right panel) and effective equation of state
$w_{\Lambda}(z)$ (left panel) of age cosmological constant with
respect to the redshift $z$, where the values
$\Omega_{\Lambda0}=0.70$, $c=10$ (solid lines), $c=15$ (dashed
lines) are adopted.}\label{fig:T}
\end{figure}

The age of the universe is defined as
\begin{equation}
t_{\Lambda}=\int^t_{0}dt'=\int^a_{0}\frac{da'}{a'H}.\label{eq:time}
\end{equation}
Taking it as the role of time scale, one has the vacuum energy
density
\begin{equation}
\rho_{\Lambda}=3c^2M^2_P/t^2_{\Lambda},\label{eq:rhot}
\end{equation}
where $c$ is the model constant. Defining the dimensionless energy
densities $\Omega_{m}=\rho_{m}/(3M^2_PH^2)$ and
$\Omega_{\Lambda}=\rho_{\Lambda}/(3M^2_PH^2)$, the Friedmann
equation is rewritten as
\begin{equation}
\Omega_{m}+\Omega_{\Lambda}=1.
\end{equation}
The energy conservation equation (\ref{eq:conservation}) can be
rewritten as
\begin{equation}
\frac{d\ln
H}{dx}+\frac{3}{2}\left(1-\Omega_{\Lambda}\right)=0,\label{eq:diffH}
\end{equation}
where $x=\ln a$. Adjoining Eq. (\ref{eq:time}), Eq. (\ref{eq:rhot})
and the definition of the dimensionless energy density
$\Omega_{\Lambda}$, one has
\begin{equation}
\int^{a}_{0}\frac{d\ln
a'}{H}=\frac{c}{H}\sqrt{\frac{1}{\Omega_{\Lambda}}}.\label{eq:integral}
\end{equation}
Taking the derivative with respect to $x=\ln a$ from the both sides
of the above equation (\ref{eq:integral}), one has the differential
equation
\begin{equation}
\frac{d\ln
H}{dx}+\frac{1}{2}\frac{d\ln\Omega_{\Lambda}}{dx}+\frac{\sqrt{\Omega_{\Lambda}}}{c}=0.
\end{equation}
Substituting Eq. (\ref{eq:diffH}) into above differential equation,
one obtains the differential equation of $\Omega_{\Lambda}$
\begin{equation}
\Omega_{\Lambda}'=\Omega_{\Lambda}\left(3-3\Omega_{\Lambda}-\frac{2}{c}\sqrt{\Omega_{\Lambda}}\right),\label{eq:EHEQ}
\end{equation}
where $'$ denotes the derivative with respect to $x=\ln a$. This
equation describes the evolution of the dimensionless energy density
of dark energy. Clearly, one can see that this equation is different
from the corresponding one derived in \cite{ref:agegraphic1}, given
the initial conditions, and the evolution of the age cosmological
constant is different from the one of agegraphic dark energy. With
the derivative of the variable of redshift $z$, the above equation
can be rewritten as
\begin{equation}
\frac{d\Omega_{\Lambda}}{dz}=-\Omega_{\Lambda}\left[3\left(1-\Omega_{\Lambda}\right)-\frac{2\sqrt{\Omega_{\Lambda}}}{c}\right](1+z)^{-1}.\label{eq:TZ}
\end{equation}
From Eq. (\ref{eq:EEOS}), it is easy to obtain the effective
equation of state of dark energy
\begin{eqnarray}
w^{eff}_{\Lambda}&=&-1-\frac{1}{3}\frac{d \ln \rho_{\Lambda}}{d\ln
a}\nonumber\\
&=&-1+\frac{2}{3}\frac{\sqrt{\Omega_{\Lambda}}}{c}.
\end{eqnarray}
It is in the range of $-1<w^{eff}_{\Lambda}<-1+2/(3c)$, when one
notices the dark energy density rasio $0\le\Omega_{\Lambda}\le1$.
The form of the effective equation of state of the horizon
cosmological constant is different from the one of agegraphic dark
energy. In the earlier time where $\Omega_{\Lambda}\approx 0$, one
has $w^{eff}_{\Lambda}\rightarrow -1$. It means at earlier time, the
conformal time cosmological constant behaves just like a true
cosmological constant. When the age cosmological constant dominates,
its property depends on the parameter $c$ strongly. One can also
easily have the deceleration parameter
\begin{eqnarray}
q&=&-\frac{\dot{H}+H^2}{H^2}\nonumber\\
&=&-1-\frac{d\ln H}{d\ln a}\nonumber\\
&=&\frac{1}{2}-\frac{3}{2}\Omega_{\Lambda}.
\end{eqnarray}
To have an current accelerated expansion of the universe,
$\Omega_{\Lambda0}>1/3$ is required. In Fig. \ref{fig:T}, the
evolution curves with respect to redshift $z$ is plotted with
different values of parameter $c$ when the initial value
$\Omega_{\Lambda0}=0.70$ is taken.

\subsection{Conformal time as time scale}\label{sec:CT}

\begin{figure}[tbh]
\centering
\includegraphics[width=5in]{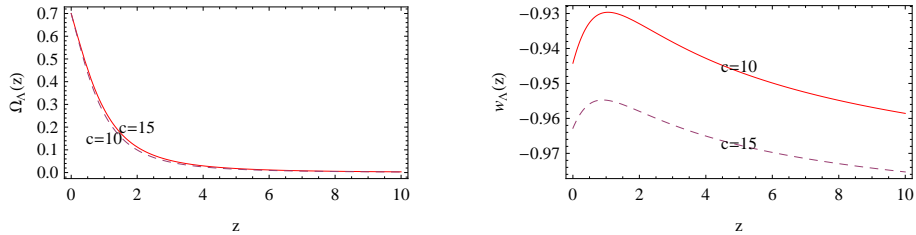}
\caption{The evolutions of dimensionless density parameter
$\Omega_{\Lambda}(z)$ (right panel) and effective equation of state
$w_{\Lambda}(z)$ (left panel) of age cosmological constant with
respect to the redshift $z$, where the values
$\Omega_{\Lambda0}=0.70$, $c=10$ (solid lines), $c=15$ (dashed
lines) are adopted.}\label{fig:CT}
\end{figure}

The conformal time is defined as
\begin{equation}
\eta_{\Lambda}=\int^{t}_{0}\frac{dt'}{a}=\int^{a}_{0}\frac{da'}{a'^{2}H}.
\end{equation}
In this case, the vacuum energy density is given as
\begin{equation}
\rho_{\Lambda}=3c^2M^2_P/\eta^2_{\Lambda}.\label{eq:rhoCT}
\end{equation}
Repeating the analysis and calculations as done in \ref{sec:ATS},
one has the differential equation of $\Omega_{\Lambda}$
\begin{equation}
\Omega_{\Lambda}'=\Omega_{\Lambda}\left[3-3\Omega_{\Lambda}-\frac{2\sqrt{\Omega_{\Lambda}}}{ca}\right],\label{eq:PHEQ}
\end{equation}
where $'$ denotes the derivative with respect to $x=\ln a$. With the
derivative of the variable of redshift $z$, the above equation can
be rewritten as
\begin{equation}
\frac{d\Omega_{\Lambda}}{dz}=-\Omega_{\Lambda}\left[3\left(1-\Omega_{\Lambda}\right)(1+z)^{-1}-\frac{2\sqrt{\Omega_{\Lambda}}}{c}\right].\label{eq:CTZ}
\end{equation}
The effective equation of state and deceleration parameter are given
as
\begin{eqnarray}
w^{eff}_{\Lambda}&=&-1+\frac{2}{3}\frac{\sqrt{\Omega_{\Lambda}}}{c}(1+z),\\
q&=&\frac{1}{2}-\frac{3}{2}\Omega_{\Lambda}.
\end{eqnarray}
These equations are different from the ones derived in
\cite{ref:agegraphic2}. It is easy to see that the behavior of
conformal time cosmological constant is rather different from that
of age cosmological constant. At later time where $a\rightarrow
\infty$, the effective equation of state goes to
$w^{eff}_{\Lambda}\rightarrow -1$, it mimics a true cosmological
constant regardless of the value of $c$. The evolution curves with
respect to redshift $z$ is plotted in Fig. \ref{fig:CT}, where the
different values of parameter $c$ and the initial value
$\Omega_{\Lambda0}=0.70$ are taken respectively.

\section{Conclusions}\label{sec:Con}

In this paper, time variable cosmological constants, dubbed {\it age
cosmological constants}, are investigated inspired by the
observations that any nonzero value of the cosmological constant
$\Lambda$ can introduce a natural length scale and time scale into
Einstein's theory. Reversely, a variable cosmological time or length
scale can introduce a time variable positive tiny cosmological
constant. Here the age of our universe and conformal time were used
as time scales. The results are rather different from that of the
so-called agegraphic dark energy models, though the effective
equation of state of age cosmological constants are common. But for
the different evolution equations of the dimensionless energy
density, please see Eq.(\ref{eq:EHEQ}) and Eq.(\ref{eq:PHEQ}) (or
Eq.(\ref{eq:TZ}) and Eq.(\ref{eq:CTZ})). So, the whole evolution
history will be different from that of agegraphic dark energy
models. In Fig. \ref{fig:T} and Fig. \ref{fig:CT}, the evolution
curves were plotted with different values of the model parameter $c$
and the same initial condition $\Omega_{\Lambda0}=0.7$. It can be
seen that the model only contains the model parameter $c$, here we
just put some values of the parameter $c$ and leave the 'precise'
value to be obtained by fitting cosmic observations such as type Ia
supernovae, CMB and BAO etc.

\acknowledgements{This work is supported by NSF (10703001), SRFDP
(20070141034) of P.R. China.}

\end{document}